\documentclass[twocolumn]{aastex631}

\usepackage{appendix}
\usepackage{amsmath}

\begin{document}
\interfootnotelinepenalty=10000
\title{A rest-frame near-IR study of clumps in galaxies at $1<z<2$ using JWST/NIRCam: connection to galaxy bulges }

\correspondingauthor{Boris S. Kalita}
\email{boris.kalita@ipmu.jp}

\author[0000-0001-9215-7053]{Boris S. Kalita}
\altaffiliation{Kavli Astrophysics Fellow}
\affiliation{Kavli Institute for the Physics and Mathematics of the Universe, The University of Tokyo, Kashiwa, 277-8583, Japan }
\affiliation{Kavli Institute for Astronomy and Astrophysics, Peking University, Beijing 100871, People{\textquotesingle}s Republic of China}
\affiliation{Center for Data-Driven Discovery, Kavli IPMU (WPI), UTIAS, The University of Tokyo, Kashiwa, Chiba 277-8583, Japan}

\author[0000-0002-0000-6977]{John D. Silverman}
\affiliation{Kavli Institute for the Physics and Mathematics of the Universe, The University of Tokyo, Kashiwa, 277-8583, Japan }
\affiliation{Department of Astronomy, School of Science, The University of Tokyo, 7-3-1 Hongo, Bunkyo, Tokyo 113-0033, Japan}
\affiliation{Center for Data-Driven Discovery, Kavli IPMU (WPI), UTIAS, The University of Tokyo, Kashiwa, Chiba 277-8583, Japan}

\author[0000-0002-3331-9590]{Emanuele Daddi}
\affiliation{CEA, Irfu, DAp, AIM, Universit\`e Paris-Saclay, Universit\`e de Paris, CNRS, F-91191 Gif-sur-Yvette, France}

\author[0000-0003-4758-4501]{Connor Bottrell}
\affiliation{Kavli Institute for the Physics and Mathematics of the Universe, The University of Tokyo, Kashiwa, 277-8583, Japan }

\author[0000-0001-6947-5846]{Luis C. Ho}
\affiliation{Kavli Institute for Astronomy and Astrophysics, Peking University, Beijing 100871, People{\textquotesingle}s Republic of China}
\affiliation{Department of Astronomy, School of Physics, Peking University, Beijing 100871, People{\textquotesingle}s Republic of China}

\author[0000-0001-8917-2148]{Xuheng Ding}
\affiliation{Kavli Institute for the Physics and Mathematics of the Universe, The University of Tokyo, Kashiwa, 277-8583, Japan }

\author[0000-0002-8434-880X]{Lilan Yang}
\affiliation{Kavli Institute for the Physics and Mathematics of the Universe, The University of Tokyo, Kashiwa, 277-8583, Japan }



\begin{abstract}
\noindent

A key question in galaxy evolution has been the importance of the apparent `clumpiness' of high redshift galaxies. Until now, this property has been primarily investigated in rest-frame UV, limiting our understanding of their relevance. Are they short-lived or are associated with more long-lived massive structures that are part of the underlying stellar disks? We use JWST/NIRCam imaging from CEERS to explore the connection between the presence of these `clumps' in a galaxy and its overall stellar morphology, in a mass-complete ($log\,M_{*}/M_{\odot} > 10.0$) sample of galaxies at $1.0 < z < 2.0$. Exploiting the uninterrupted access to rest-frame optical and near-IR light, we simultaneously map the clumps in galactic disks across our wavelength coverage, along with measuring the distribution of stars among their bulges and disks. Firstly, we find that the clumps  are not limited to rest-frame UV and optical, but are also apparent in near-IR with $\sim 60\,\%$ spatial overlap. This rest-frame near-IR detection indicates that clumps would also feature in the stellar-mass distribution of the galaxy. A secondary consequence is that these will hence be expected to increase the dynamical friction within galactic disks leading to gas inflow. We find a strong negative correlation between how clumpy a galaxy is and strength of the bulge. This firmly suggests an evolutionary connection, either through clumps driving bulge growth, or the bulge stabilizing the galaxy against clump formation, or a combination of the two. Finally, we find evidence of this correlation differing from rest-frame optical to near-IR, which could suggest a combination of varying formation modes for the clumps.


\end{abstract}

\keywords{galaxies: evolution – galaxies: formation – galaxies: structure}


\section{Introduction} \label{sec:intro}




Over the last two decades, extremely deep high resolution data, courtesy mainly of the Hubble Space telescope, revealed high redshift star-forming galaxies to be much more clumpy than their low redshift counterparts \citep{conselice04,elmegreen05,elmegreen08,forster11,guo15,guo18,shibuya16}. This has sparked a debate about the consequences of these structures on galaxy evolution. Mainly observed in rest-frame UV and optical wavelengths, the origin and evolution of these `clumps' within star-forming galaxies are far from well understood. 

Properties such as stellar-masses, stellar-ages, sizes and gas (or dust) content have been investigated in observational studies \citep{elmegreen07, guo12, guo18, soto17, zanella19, rujopakarn23, martin23, sattari23} of these kpc-scale star-forming complexes \citep[resembling `localized star-bursts';][]{wuyts12, wuyts13, bournaud15, zanella15, mieda16}. In parallel, simulations have been driving forward our theoretical understanding of their origin, evolution and subsequent fate \citep[e.g.,][]{elmegreen08,dekel09, bournaud09, ceverino10, ceverino12, mandelker17}.

One of the most consequential questions, concerning the morphological evolution of galaxies, is whether these are simply very low-mass features only observed in rest-frame UV and optical, or massive enough to contribute to the total stellar-mass tracing rest-frame near-IR light. Some simulations suggest that they are disrupted on short timescales ($\lesssim 50\,\rm Myr$), thus having minimal effect on the underlying stellar distribution \citep{murray10,tamburello15,buck17,oklopvic17}. However, other simulations suggest that clumps survive for much longer and are expected to be crucial drivers of bulge growth through dynamical friction and gravitational torques \citep{bournaud11, bournaud14, elmegreen08, ceverino10, mandelker14, mandelker17}. They could also be associated with more massive structures that HST-based rest-frame UV observations miss \citep{faure21}. 


Before the James Webb Space Telescope (JWST) era, observations were limited to star-formation tracing rest-frame UV and optical light, mainly using HST \citep[e.g.,][]{guo15, guo18, sattari23}. We did not have high resolution capabilities in rest-frame near-IR at $z > 1$ to map the underlying stellar distribution on the relevant spatial scales ($\sim 1\,\rm kpc$). With the highly sensitive and high resolution rest-frame optical (better than HST) and near-IR  capacities of JWST, we can finally undertake a relatively more complete study. It is now possible to extend the resolved imaging of these clumpy galaxies well into rest-frame near-IR, thereby accessing the bulk of the stellar light. 


In this work, we use the capabilities of JWST/NIRCam to make the first resolved maps of clumps within galaxies in both the rest-frame optical and near-IR for galaxies at $z > 1$. It should be noted that the term `clumps' has until now been used to refer to marginally resolved or un-resolved structures and are expected to be a result of gravitational instabilities. These conclusions are based on in-depth investigations conducted primarily in rest-frame UV. The structures detected in this work are at rest-frame optical and near-IR, and albeit we expect them to be associated with the previously studied UV-detected clumps (Sec.~\ref{sec:disk_frag_stats}), the exact connection will be tackled in a follow-up paper (Kalita et. al. in prep). Nevertheless, we still use the same term `clumps' in this work as a general reference to structures giving galaxies a clumpy appearance. We are not implying that they necessarily share the same properties as the structures previously studied.



To make a parallel assessment of the general stellar morphology of each galaxy, we exploit the longest wavelength NIRCam band (F444W) to obtain a bulge-to-disk flux ratio. This value closely traces the corresponding stellar-mass distribution due to a minimally varying mass-to-light ratio at such high a wavelength \citep{bell03, zibetti09, schombert19}. Throughout, we adopt a concordance $\Lambda$CDM cosmology, characterized by  $\Omega_{m}=0.3$, $\Omega_{\Lambda}=0.7$, and $\rm H_{0}=70$ km s$^{-1}\rm Mpc^{-1}$. We use a Chabrier initial mass function. All images are oriented such that north is up and east is left.

\section{Sample selection}


We use of the Cosmic Evolution and Epoch of Reionization Survey (CEERS; ERS 1345, PI: S Finkelstein), a JWST Early Release Science programs. CEERS observed a section of the Extended Goth Strip Hubble Space Telescope (EGS-HST) field using NIRCam Wide-band imaging. The reduced images \citep[made available by the CEERS collaboration;][]{bagley23} span a large wavelength window from $1\,\mu$m to $5\,\mu$m, covered by six filters (with average $5\,\sigma$ depths): F115W (29.1 mag), F150W (29.0 mag), F200W (29.2 mag), F277W (29.2 mag), F356W (29.2 mag) and F444W (28.6 mag)\footnote{We do not incorporate the F410M data into this study due to it being shallower by a factor of $\sim 2$ than the rest of the data. F444W also has a similar issue, due to which we shall be excluding measurements of clumpiness in this filter later. }.

We require simultaneous coverage of rest-frame optical as well as near-IR light, along with similar physical to angular scales (to not introduce systematic biases) across our sample. Moreover, the galaxies should preferentially be at $z > 0.5$, where most studies in literature begin finding a significant population of clumpy galaxies \citep{guo15}. To satisfy each of these requirements, we settle on a redshift range of $1.0 < z < 2.0$.

For the sample selection itself, we use the catalogue for the EGS-HST field \citep[][ S17]{stefanon17} created with an extremely large wavelength coverage ($0.4 - 8.0\,\mu$m). Its stellar-mass completeness including a maximum dust attenuation value, $A_{v} = 3.0\,$mag, is set at $log (M_{*}/M_{\odot}) = 9.5$ for $1.0 < z < 1.5$ and $log (M_{*}/M_{\odot}) = 10.0$ for $1.5 < z < 2.0$. For the sake of convenience, we limit our analysis to galaxies above  $log (M_{*}/M_{\odot}) = 10.0$ over the whole redshift range of our sample. This method leaves us with a total of  galaxies 412 galaxies. However, we do provide the results for galaxies with $log (M_{*}/M_{\odot}) = 9.5-10.0$ at $1.0 < z < 1.5$ in the Appendix.

\section{Methodology}
\subsection{Stellar morphology: bulges and disks} \label{subsec: stell_morph}
 We begin with an assessment of the stellar light variation by measuring the bulge-to-disk flux ratio of each galaxy. To ensure that this reflects the underlying stellar-mass distribution, we use the longest wavelength NIRCam filter available to us, F444W. For our redshift range, we expect a corresponding variation of mass-to-light ratio $\lesssim 25 - 30\%$ \citep{schombert19, zibetti09, bell03} with color. This would have increased had we used shorter wavelengths, thereby complicating the interpretation of our results. Therefore, this decision to use the longest wavelengths is aimed at using an `almost' color-invariant tracer of the stellar distribution. An additional advantage of this approach is the sensitivity to highly obscured cores/bulges that we expect to find at high-z \citep[e.g.,][]{kalita22}, since attenuation is minimal\footnote{to quote a generous upper limit, assuming an $\rm A_{V} = 3.0$, we get an attenuation of 0.85 in F444W for a typical \cite{cardelli89} Milky Way extinction curve} in rest-frame near-IR.
 
We measure the flux of the bulge and the rest of the galaxy by fitting our sample with a dual component model: two S\'{e}rsic profiles with fixed indices of $n = 1$ for the disk and $n = 4$ for the bulge. While we use this parameter settings to obtain the results presented throughout this work, we use also redo our measurements with a S\'{e}rsic index of $2$ \citep[associated with a pseudo-bulge;][]{gadotti09} in place of $4$ for the (classical) bulge and find only negligible differences (well within the uncertainties). Moreover, the use of $n = 4$ is driven by our choice to not obtain a perfect fit on an individual galaxy level but rather achieve a uniform determination of the flux across our sample \citep[this being a widely used approach, e.g.,][]{simard11, meert11, bottrell17a, bottrell17b, bottrell19}. Finally, it also allows us to broadly separate the disk and bulge.

For each object, we make cutouts of dimensions $101 \times 101$ pixels ($3^{\prime\prime} \times 3^{\prime\prime}$) of the background subtracted F444W images. We also create corresponding cutouts of the available weight maps from CEERS, that is used as the noise maps in the fitting procedures. Finally, the PSF for the fitting is generated by median-stacking 7 unsaturated stars we found within the field-of-view of the observations. The fitting is done using a python-based package GALIGHT \footnote{https://github.com/dartoon/galight} \citep{ding22}, which in turn implements the forward-modelling galaxy image fitting tool LENSTRONOMY \citep{birrer18}. The approach allows us access to the full posterior distribution of each fitted parameter. This fitting is then optimised through a two-stage Particle Swarm Optimizer \citep[PSO]{kennedy95}, which is finally fed into a Markov Chain Monte Carlo (MCMC) fitting as the initial condition. We are then left with the best-fit parameter as well as their respective $1 \sigma$ confidence intervals. 

Finally, we obtain three morphological flux ratios from this analysis: the bulge-to-disk (using the fluxes of the two components), bulge-to-total (where the total flux is given by the sum of the disk and residual flux) and disk-to-total. We will mainly be using the bulge-to-disk ratio throughout this work, but will refer to the other two to in order to further our understanding of the results (in Sec.~\ref{sec:bdr_frag_corr})  
\subsection{Quantifying the clumpiness} \label{sec:disk-fragments}

\begin{figure*}[!ht]
    \centering
    \includegraphics[width=0.95\textwidth]{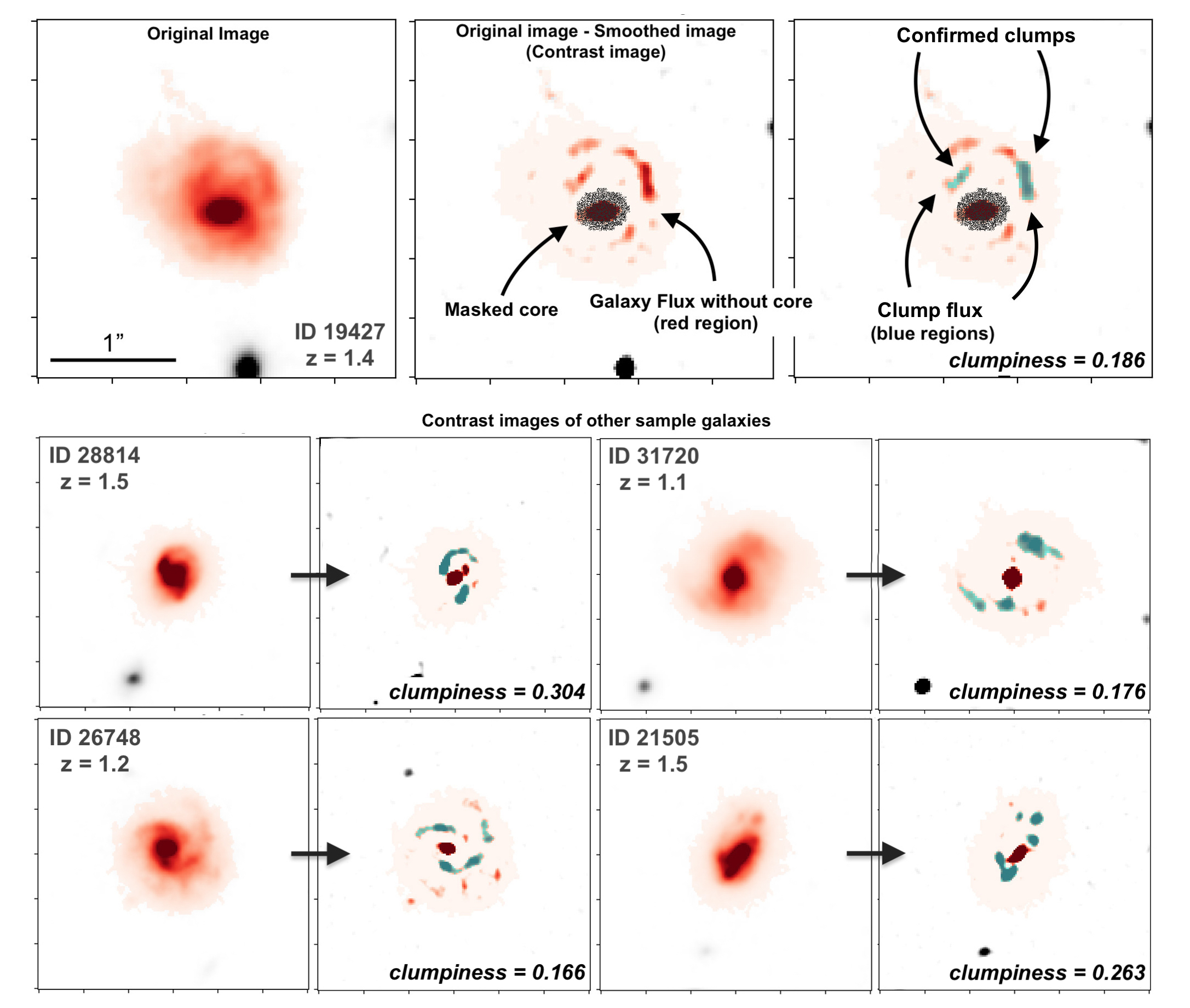}
    \caption{The clump detection: (top-left) An F115W image of a galaxy within our sample (with intermediate levels of clumpiness) with an equivalent PSF as that for F444W. The $2\,\sigma$ segmentation map for the galaxy at F444 has a red color-map to distinguish it from other sources. (Top-middle) This image is smoothed using a Gaussian filter of $\sigma = 4$\,pixels and then subtracted from the original image to get the contrast image. The core is masked during this detection algorithm and is shown with an overlaid blackened mesh. The remaining `red region' gives the net galaxy flux without the bulge. (Top-right) The source detection algorithm discussed in Sec. \ref{sec:disk-fragments}  results in the collection of confirmed clumps  highlighted in blue. The net flux in these regions is divided by the disk flux to get the clumpiness. (Middle and bottom) The lower panels show F115W images and their corresponding contrast images (with highlighted regions of clumps) of other objects within our sample. The S17 IDs for each object is also provided.}
    \label{fig:disk-fragment_detection}
\end{figure*}

\begin{figure}[!ht]
    \centering
    \includegraphics[width=0.45\textwidth]{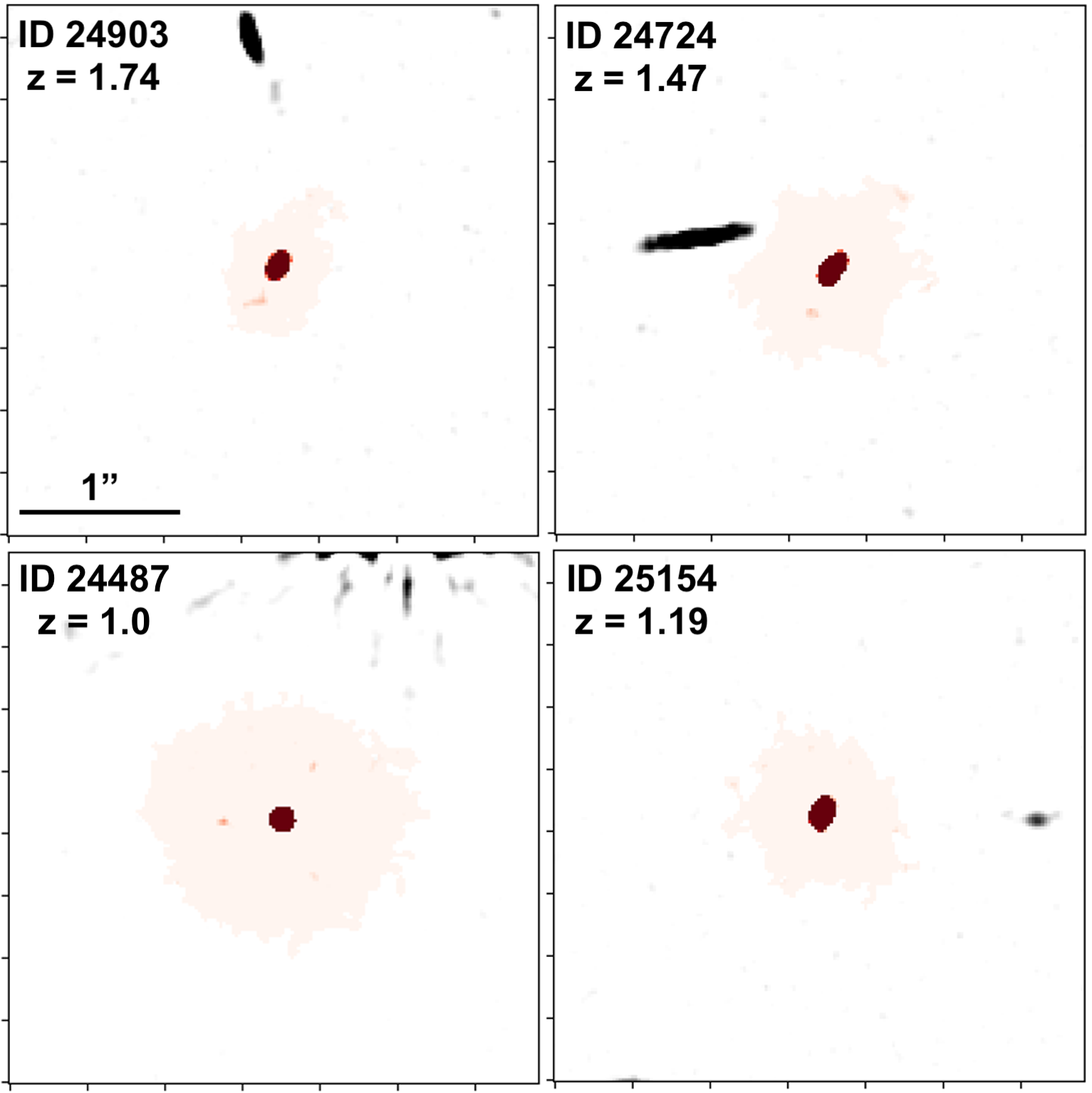}
    \caption{Contrast images similar to those in Fig.~\ref{fig:disk-fragment_detection}, but of galaxies which had no clumps detected.}
    \label{fig:disk-fragment_non_detection}
\end{figure}

\begin{figure*}[!ht]
    \centering
    \includegraphics[width=0.90\textwidth]{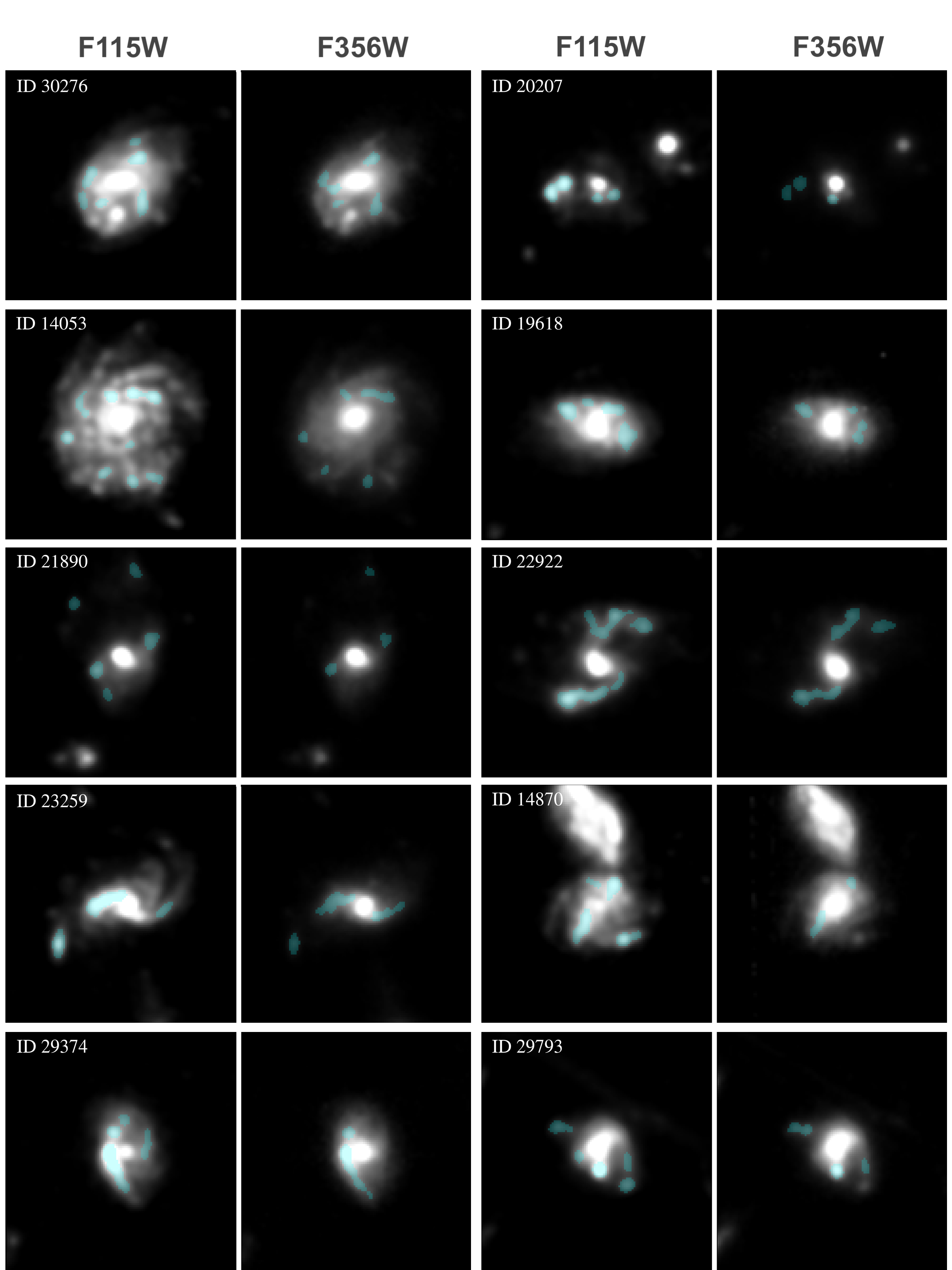}
    \caption{Examples of galaxies with detected clumps. For each, the rest-frame optical (F115W) and rest-frame near-IR (F356W) images are shown. The detected clumps using the method shown in Fig.\ref{fig:disk-fragment_detection} are shown with a blue color scheme. It should be noted that the the clumps may not always be clearly detected by eye on these original images, for which one would rather require the respective contrast maps.}
    \label{fig:disk-fragment_detection_examples}
\end{figure*}
\begin{table*}
\centering
\caption{Stellar-mass complete ($10.0 < log (M_{*}/M_{\odot}) < 11.0$) sample used in this work. 
\label{tab:sample_stats}}
\begin{tabular}{lcccc}
 & Total Galaxies  & Bulge+disk measurements &  clumpiness   detected \\
 &  &  & ($> 68\,\%\, \rm confidence$)\\
 &  &  & F115W\,\,\,F150W\,\,\,F200W\,\,\,F277W\,\,\,F356W \\\hline \\
 $1.0 < z < 1.5$ ($> 10^{10.0}\,M_{\odot}$) & 213 & 156 & 99\,\,\,\,\,\,\,\,\,\,\,\,\,\,\,94\,\,\,\,\,\,\,\,\,\,\,\,\,\,\,86\,\,\,\,\,\,\,\,\,\,\,\,\,\,\,88\,\,\,\,\,\,\,\,\,\,\,\,\,\,\,82\\
 $1.0 < z < 1.5$ ($10^{10.0}-10^{10.5}\,M_{\odot}$) & 114 & 79 & 48\,\,\,\,\,\,\,\,\,\,\,\,\,\,\,48\,\,\,\,\,\,\,\,\,\,\,\,\,\,\,45\,\,\,\,\,\,\,\,\,\,\,\,\,\,\,42\,\,\,\,\,\,\,\,\,\,\,\,\,\,\,41\\
 $1.0 < z < 1.5$ ($10^{10.5}-10^{11.0}\,M_{\odot}$) & 99 & 77 & 51\,\,\,\,\,\,\,\,\,\,\,\,\,\,\,46\,\,\,\,\,\,\,\,\,\,\,\,\,\,\,41\,\,\,\,\,\,\,\,\,\,\,\,\,\,\,46\,\,\,\,\,\,\,\,\,\,\,\,\,\,\,41\\ \\
 \hline \\
 $1.5 < z < 2.0$ ($> 10^{10.0}\,M_{\odot}$) & 199 & 136 & 86\,\,\,\,\,\,\,\,\,\,\,\,\,\,\,83\,\,\,\,\,\,\,\,\,\,\,\,\,\,\,81\,\,\,\,\,\,\,\,\,\,\,\,\,\,\,77\,\,\,\,\,\,\,\,\,\,\,\,\,\,\,69\\
 $1.5 < z < 2.0$ ($10^{10.0}-10^{10.5}\,M_{\odot}$) & 101 & 69 & 41\,\,\,\,\,\,\,\,\,\,\,\,\,\,\,38\,\,\,\,\,\,\,\,\,\,\,\,\,\,\,37\,\,\,\,\,\,\,\,\,\,\,\,\,\,\,36\,\,\,\,\,\,\,\,\,\,\,\,\,\,\,34\\
 $1.5 < z < 2.0$ ($10^{10.5}-10^{11.0}\,M_{\odot}$) & 98 & 67 & 45\,\,\,\,\,\,\,\,\,\,\,\,\,\,\,45\,\,\,\,\,\,\,\,\,\,\,\,\,\,\,44\,\,\,\,\,\,\,\,\,\,\,\,\,\,\,41\,\,\,\,\,\,\,\,\,\,\,\,\,\,\,35\\
 
\\ \hline
\end{tabular}
\end{table*}

\begin{figure}[!ht]
    \centering
    \includegraphics[width=0.5\textwidth]{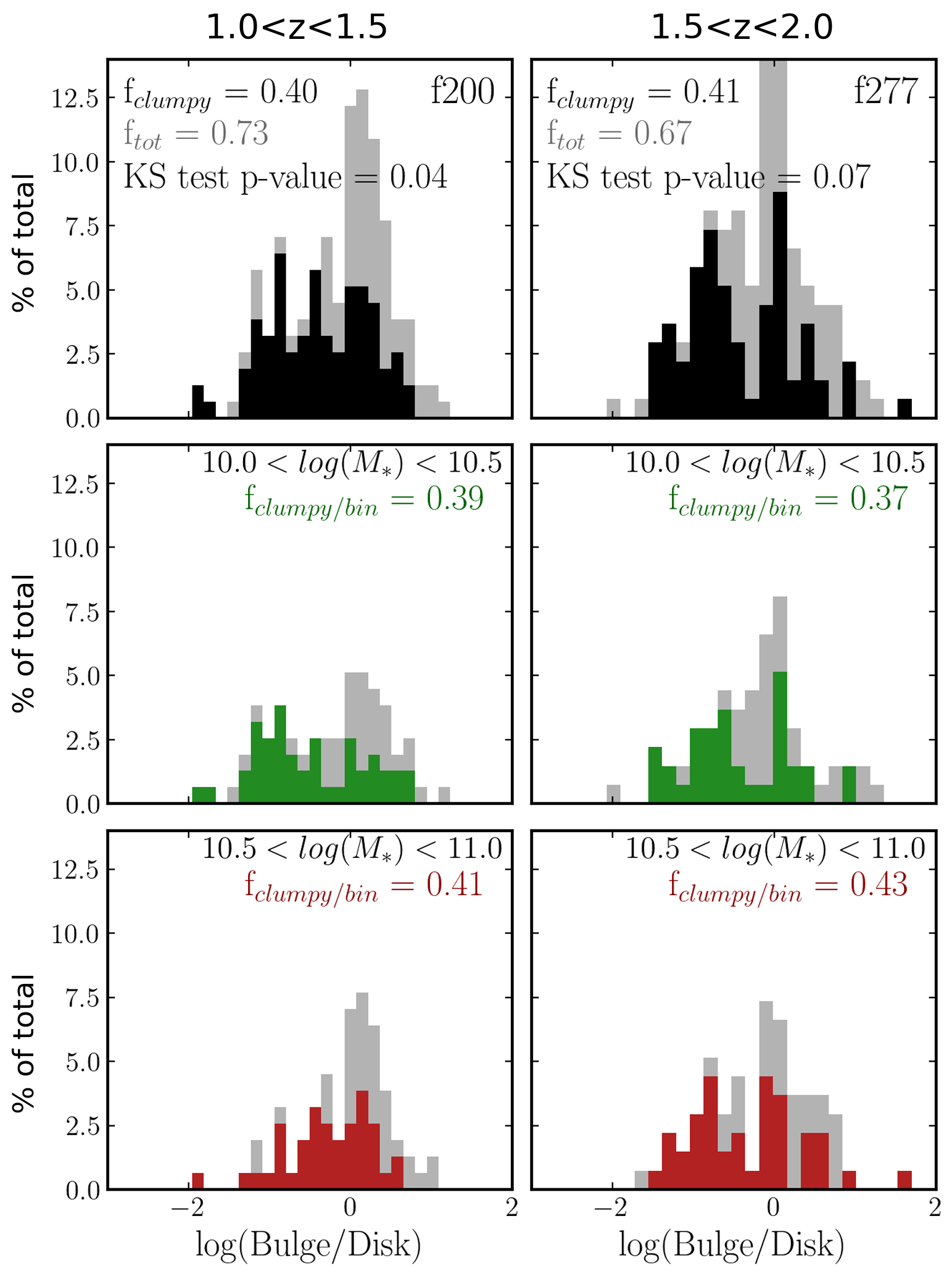}
    \caption{(Top) Histogram showing the bulge-to-disk ratio distribution of all galaxies that have both significant bulge and disk flux measurements. (in grey, which makes up a fraction `f$_{\rm tot}$' of all galaxies in S17 within our mass and redshift brackets). The black histogram is made up of the subset of sources that have clumpiness detected at rest-frame wavelength $\sim 1\,\mu$m (f200W and f277W for the two redshift windows), which makes up a fraction `f$_{\rm clumpy}$' of all the S17 galaxies. The probabilty of the f$_{\rm tot}$ and `f$_{\rm clumpy}$' samples being from the same distribution (the p-value) is also shown. (Middle and bottom) The same histogram but limited to the stellar-mass ranges $log (M_{*}/M_{\odot}) = 10.0-10.5$ and $log (M_{*}/M_{\odot}) = 10.0-11.0$ respectively. The `f$_{\rm clumpy/bin}$' is the `f$_{\rm clumpy}$' counted for the individual stellar-mass windows. The two columns correspond to the redshift ranges $1.0-1.5$ and $1.5-2.0$.}
    \label{fig:hist_1}
\end{figure}
The aim of the second segment of our analysis is to quantify the flux from clumps  that one can observe by visual inspection in the highly sensitive JWST/NIRCam images of the sample galaxies. As we shall be making comparisons across the wide wavelength coverage of the data (from F115W to F444W), we begin by matching the PSF sizes to that of the longest wavelength and therefore the lowest resolution filter (F444W). We obtain the PSFs of each filter as discussed in Sec. \ref{subsec: stell_morph}. A Gaussian-model fitting gives us the PSF parameters which are used to calculate the effective $\sigma$ of the Gaussian kernel that each image needs to be convolved with to match the resolution of the F444W data.  

We build an automated clump detection algorithm (Fig. \ref{fig:disk-fragment_detection}) based on the method widely used for detecting clumps in galaxies in HST UV and optical bands \citep{conselice03, guo15, calabro19}. We first smooth the measurement image (from F115W to F444W, but with a PSF equivalent to that of F444W), using a Gaussian filter of $\sigma = 4$\,pixels.\footnote{ We try a range of values for $\sigma = 3 - 6$\,pixels, a range based on the values used in previous works. Albeit the results of this paper do not change beyond the already estimated errors, we use $\sigma = 4$\,pixels as it picks the finest clumps  we detect through a visual inspection of the JWST images.} This smoothed image is subtracted from the original image, leaving behind a contrast map showing structures varying at scales similar to the difference of the F444W PSF and the convolution of it with the Gaussian used for smoothing. This difference is found to be $\sim 0.07^{\prime\prime}$. It should be noted that this is not the size of the clumps. We then make cutouts from the measurement images corresponding to each source of dimensions $3^{\prime\prime} \times 3^{\prime\prime}$ as done in the previous section.    

We then run a source detection using the python package PHOTUTILS \citep{photutils}, also included in GALIGHT, with a threshold\footnote{There are two reasons we use a $5\,\sigma$ detection threshold with PHOTUTILS: Using thresholds that are lower results in large sections of the galaxies rather the clumps.   Moreover, the robustness check carried out for each detection leads to the rejection of more than $\gtrsim 60 \%$ of the sources at $3\,\sigma$ and $\gtrsim 45\%$ at $4\,\sigma$. At $5\,\sigma$ however, this rejection rate is around 20-30\%. Above this threshold our sample starts shrinking considerably. Hence we settle at the $5\,\sigma$ threshold.} at $5\,\sigma$/pixel. This method selects regions with peaked emission associated with clumps  within the galaxies (Fig.~\ref{fig:disk-fragment_detection}). We only consider clumps  within the extent of each galaxy, which is determined by a source detection on the stellar-mass sensitive F444W image with a threshold of $2\,\sigma$ (the red color map region in Fig.~\ref{fig:disk-fragment_detection}). Moreover, it is critical to not include the central bulge as a clump since both are interpreted as small-scale structures by our algorithm. Hence, we mask the central region\footnote{The central mask is determined by the extent of the central clump  (using the contrast map) as seen in the segmentation map for F444W. The central position is determined by finding the point of minimum asymmetry in the F444W band, where the central bulge is the most dominant object. This is always found to be coincident with the bulge location in the bulge-disk decomposition fit.} as shown in the top panel of Fig.~\ref{fig:disk-fragment_detection}. 

To ensure that our detection is robust, we carry out an additional test. For every single clump (associated with a continuous segmentation map source detected at $5\,\sigma$), we artificially replicate it at another random position (by replacing the flux originally there) within the galaxy image while ensuring that it is not positioned within the core or outside the galaxy edge. Then we re-run our clump detection algorithm and check if this new (fake) clump is detected. We repeat this process a 1000 times, and only allow a clump to be included in our final measurement for the galaxy if it is detected $> 68 \%$ times. This method removes $\sim 20\%$ of the original detections. The reason for the rejection includes the clumps being marginally detected, and not leading to a strong enough fluctuation within the underlying disks to be detected repeatedly. Finally, once we have the robustly detected clump map, we simply measure the net flux within them and divide it by the flux of the whole galaxy in the same filter (the region for which uses the previously used $2\,\sigma$ threshold map from the F444W image, shown in red color map in Fig.~\ref{fig:disk-fragment_detection}), after masking the core.\footnote{The reason we remove the core for the net flux measurement, is to remove any possible correlation between the clump flux fraction and the bulge measurements in Sec. \ref{subsec: stell_morph}. This method is different from how `clumpiness' is measured in most previous works, where they do not remove it. We do check however if our results change if we include the core, and we find that they do not.} This gives us the fractional flux contained in clumps, with respect to the rest of the galaxy without the core. We define this ratio as the clumpiness. 
\begin{equation*}
    \rm Clumpiness = \frac{\Sigma\,\rm Clump\,\,flux}{\rm Galaxy\,\,Flux\,\,without\,\,core}
\end{equation*}

The uncertainty in the clumpiness is determined from the error in estimation of the flux of each detection during the artificial replication process. In cases where no clumps  are detected, we add a point source within the galaxy with varying levels of fractional flux ($-4.0$ to $0.0$ in log scale) compared to the net flux of the galaxy. For each fractional flux value, we repeat the process a 1000 times. This step allows us to estimate an upper limit based on the flux level below which we begin having no robust detections ($< 68\%$). 

This process is repeated for each filter separately. Some of the results for F115W and F356W, approximately representing rest-frame optical and near-IR are shown in Fig.~\ref{fig:disk-fragment_detection_examples}.

\section{Results}
\subsection{Clumps in optical and near-IR} \label{sec:disk_frag_stats}

We focus our study to galaxies that have detectable bulges and disks (i.e. those having robust bulge and disk measurements with $< 50 \%$ uncertainty in both components). This choice removes any galaxy that is composed exclusively of a disk or of a spheroidal bulge\footnote{Note that all clumpy galaxies are retained within our sample in spite of this, as discussed in Sec.~\ref{sec:bdr_frag_corr}}. We also ensure a disk axis-ratio $> 0.3$. The latter condition is applied to prevent any biases due to high dust column density in a galaxy observed perfectly edge-on.  Our selected sample constitutes $73 \%$ and $67 \%$ of the S17 galaxies ($10.0 < log (M_{*}/M_{\odot}) < 11.0$) at redshift ranges $1.0 - 1.5$ and $1.5 - 2.0$ respectively (also evident from Table~\ref{tab:sample_stats} and f$_{tot}$ in Fig.~\ref{fig:hist_1}). This is in line with previous expectation of disk-galaxy fractions in massive galaxy samples up to $z = 2$ \citep{vanderwel14}\footnote{To check if we are indeed studying almost all disk galaxies appearing at these redshifts, we make a selection of all S17 galaxies within 0.3 dex of the star-forming main sequence and above $log (M_{*}/M_{\odot}) > 10.5$. We find that the galaxies we select, albeit after removing the disk axis-ratio constraint, make up $\gtrsim 80\,\% $ of the sample, in line with expectations for massive star-forming galaxies up to $z = 2$ in \cite{vanderwel14}.}.

At $z > 1$, studies have usually concentrated on studying clumpiness in UV bands \citep[below the $4000 \rm \AA$ break;][]{guo15}. We however concentrate on the rest-frame optical and near-IR bands to study the underlying stellar distribution, as all the JWST/NIRCam bands we use in our study are above the $4000 \rm \AA$ break (barring F115W for $1.5 < z < 2.0$, the results for which we separately discuss later). 

We first present the results at rest-frame $\sim 1\,\mu$m as representative of the whole wavelength range covered in our study. We find that out of all galaxies in our study, $40\,\%$ and $41\,\%$ of them at redshift ranges $1.0 - 1.5$ (using F200W) and $1.5 - 2.0$ (F277W) respectively show at least some level of clumpiness within our detection limits. This feature is represented as a fraction $f_{\rm clumpy}$ in Fig.~\ref{fig:hist_1}. Our fractions are of course dependent on the detection threshold we set, and using a lower value might result in a higher fraction. Nevertheless as mentioned earlier, doing so gives us less reliable clumpiness. Moreover, this somewhat strict detection still gives us results in agreement with the expected percentage of galaxies featuring UV-clumps \citep{guo15} at the respective mass and redshift ranges. Hence our conditions do not hinder the scientific relevance of our sample within the current literature framework.

As can be deduced from Table~\ref{tab:sample_stats}, similar fraction of clumpy  galaxies can be obtained across all filters spanning rest-frame optical and near-IR. We do note however that there is a drastic fall in the numbers for the F444W filter (not shown in Table~\ref{tab:sample_stats}). But it is most likely due to this being a factor $\sim 2$ shallower than the others, in addition to the clumps being fainter at long wavelengths. We therefore drop this from our clumpiness estimations and limit our analysis up to F356W, still maintaining coverage of exclusively rest-frame near-IR light in at least one filter across our redshift range. 

The similarity of fraction of galaxies showing clumpiness   across bands do not however inform us whether the same structures are being detected in each galaxy.  Thus, we apply a method of quantification of the overlap between the clumpiness  maps in rest frame optical and near-IR bands. We do so by finding the net area of all pixels that are detected in both the shortest (F115W) and longest (F356W) wavelength band used in our detection algorithm. This value is then normalised by the area of detected clumps in either of the two bands to give a percentage of associations for each. We find a value of $0.6 \pm 0.2$ and $0.7 \pm 0.2$ for rest-frame optical (F115W) and near-IR (F356W). This result strongly suggests that we are not looking at mutually exclusive clumps in separate bands.

Our results indicate that a large fraction of the galaxies showing clumpiness  in the optical and near-IR bands would also show UV-clumps, given the similarity of the fraction of galaxies showing clumpiness in our study and that in \cite{guo15}. However, we do not attempt to make a comparison to ancillary HST/acs CANDELS data due to the images being a factor of $\sim 2$ shallower than the JWST/NIRCam images we are using (F115W to F356W). Nevertheless, we observe that UV-clumps do appear in the F115W filter at $1.5 < z < 2.0$, which traces UV light at this redshift range, for all galaxies showing optical and near-IR clumps.  

Finally, a Kolmogorov-Smirnov test on the complete sample of galaxies and those with non-zero clumpiness gives only a $4\,\%$ and $7\,\%$ cumulative probablilty (p-value) of being from the same distribution. However, if we remove the clumpy galaxies from the full sample and remeasure the probabilty, we obtain p-values well below $1\,\%$. We therefore interpret this result as an indication of clumpy galaxies being largely independent of galaxies with no clumps, in regards to their bulge-to-disk ratios.


\subsection{Bulge-to-disk ratio and clumpiness} \label{sec:bdr_frag_corr}

\begin{figure}[!ht]
    \centering
    \includegraphics[width=0.43\textwidth]{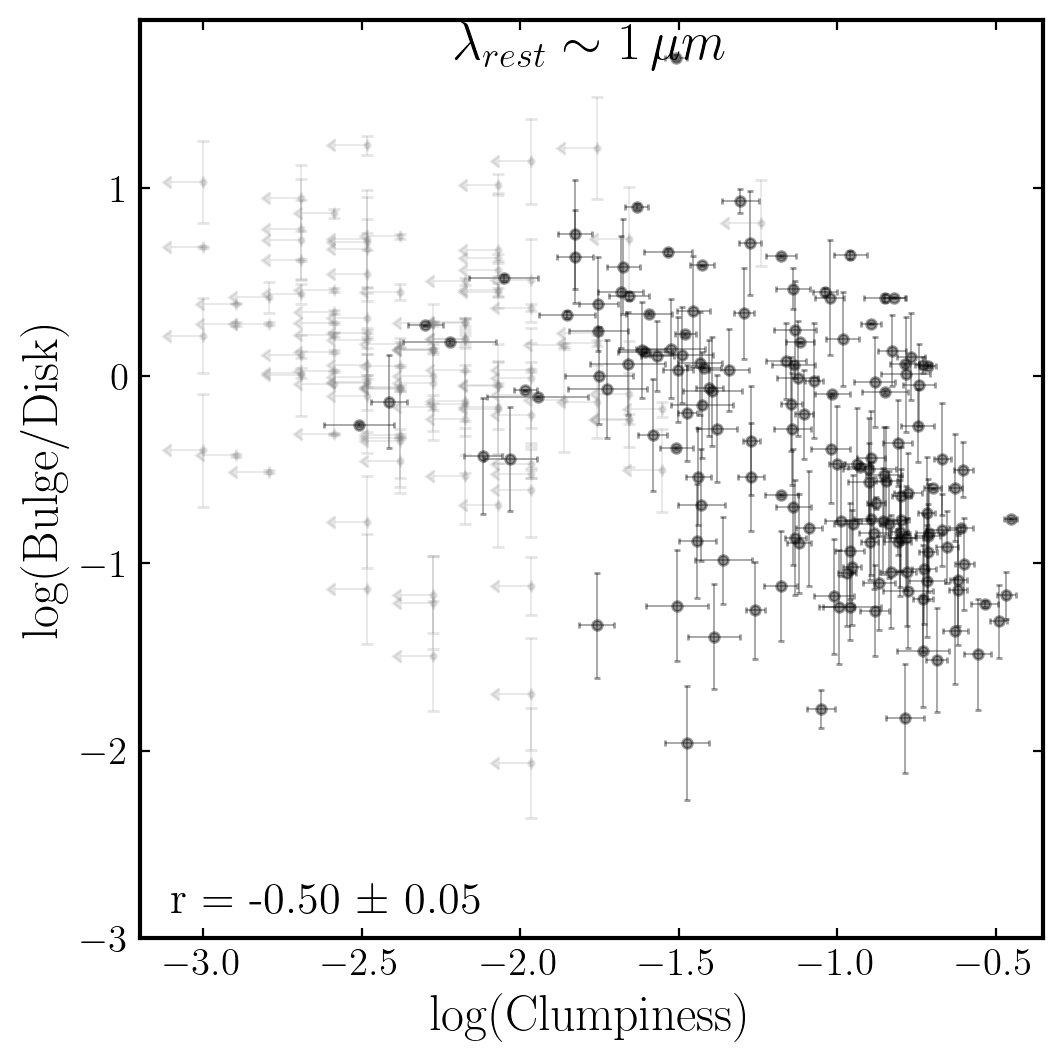}
    \caption{Bulge-to-disk ratio vs clumpiness at rest-frame wavelength $\sim 1\,\mu$m for the complete sample. For the redshift window $1.0-1.5$ we use F200, while for $1.5-2.0$ we use F277. The Pearson correlation coefficient ($r$) for the plotted data is provided in the bottom-right corner. The grey points mark the upper-limit values for all sources without clumpiness but have robust bulge-to-disk ratio. These are not included in the estimation of $r$.}
    \label{fig:bdr_frag_tot}
\end{figure}

\begin{figure}[!ht]
    \centering
    \includegraphics[width=0.45\textwidth]{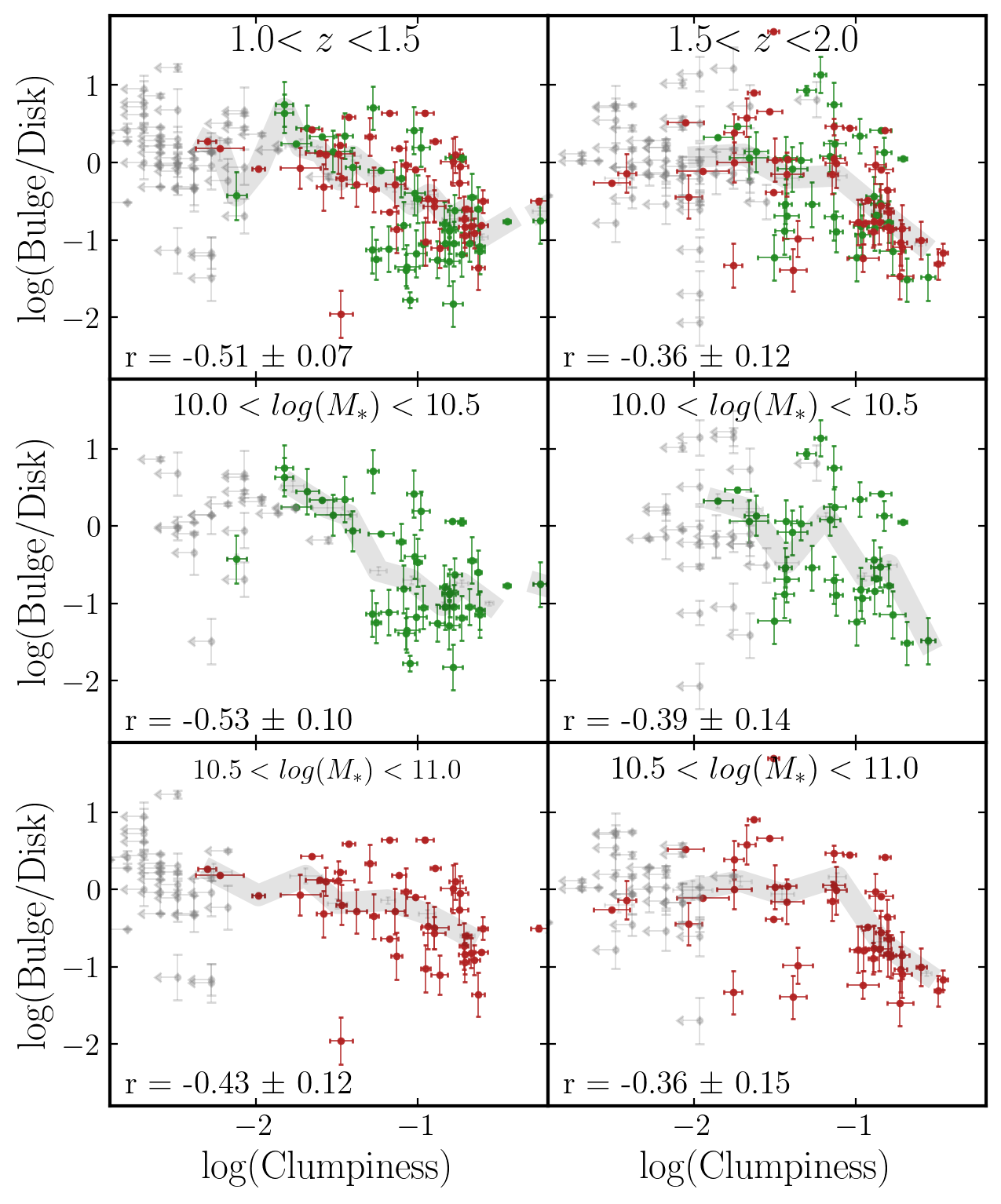}
    \caption{Bulge-to-disk ratio vs clumpiness at rest-frame wavelength $\sim 1\,\mu$m from Fig.\ref{fig:bdr_frag_tot}, split into two redshift windows $1.0-1.5$ and $1.5-2.0$. The three rows from top to bottom are the same as in Fig.~\ref{fig:hist_1}, which correspond to the full mass range (above the mass-completeness limit of $log (M_{*}/M_{\odot}) = 10.0$), $log (M_{*}/M_{\odot}) = 10.0-10.5$ and $log (M_{*}/M_{\odot}) = 10.5-11.0$ respectively. The thick grey line represents the average of the data-points and the Pearson correlation coefficient ($r$) for the plotted data is provided in the bottom-right corner in each panel. Finally, the grey points mark the upper-limit values for all sources without clumpiness but have robust bulge-to-disk ratio. }
    \label{fig:bdr_frag_1}
\end{figure}

This work exploits the ability to simultaneously map the clumps  in rest-frame optical and near-IR, along with the near-IR bulge-to-disk flux ratio (that closely traces the stellar-mass ratio between these two components as discussed in Sec.~\ref{subsec: stell_morph}). As done in Sec.~\ref{sec:disk_frag_stats}, we only show the relation between these two components at $\sim 1\,\mu$m rest-frame as a representation of our sample (Fig.~\ref{fig:bdr_frag_tot} and \ref{fig:bdr_frag_1}). For all other filters, which are found to corroborate our conclusions, please refer to the Appendix.   

Firstly, all galaxies with non-zero clumpiness values appear within our sample with robust bulge and disk measurements. Fig.~\ref{fig:bdr_frag_tot} reveals clumpiness to decrease with increasing bulge-to-disk flux ratio. This negative correlation, as determined by the Pearson correlation coefficient\footnote{The uncertainty for the coefficient is measured through bootstrapping. We randomize the pairs of values for clumpiness and bulge-to-disk ratio, and re-measure the coefficient. This process is done a total of 1000 times and the standard deviation of the distribution of the coefficient is used as the uncertainty.}, is found to be moderate to strong, with a value of $-0.50 \pm 0.05$ for the whole sample. This strength of correlation is also found to mostly hold if the sample is divided into two bins in redshift as well as stellar-mass (Fig.~\ref{fig:bdr_frag_1} along with Figs.~\ref{fig:8} and \ref{fig:9} in the appendix), with a few exceptions likely due to low number statistics.  We also note that the position of the clumpiness upper limits (in both Fig.\ref{fig:bdr_frag_tot} and ~\ref{fig:bdr_frag_1}) follow the same trend, with almost all of them appearing at the high bulge-to-disk end of the distribution. Finally, there does not seem to be any obvious evolution of this negative correlation with redshift in our sample within the current levels of uncertainty. 

We inspect if this trend is induced by the measurement method rather than an intrinsic correlation. As discussed in Sec.~\ref{subsec: stell_morph}, we already ensure that the bulge flux is not incorporated in any way in the measurement of the clumpiness \footnote{To further check this empirically, we measure the upper limits of disk clumpiness   even in galaxies where we have non-zero clumpiness values. This is done by following the same procedure discussed in Sec.~\ref{sec:disk-fragments}, but by masking the currently present clumps.   We find that there is no correlation of this limit on the bulge-to-disk ratio}. We do find that using the net-flux of the galaxy to normalise the clump flux, we still end up with a similar correlation. We also ensure that the disk axis-ratio (above the cut-off of 0.3) is not correlated to either of the two plotted parameters. Making the axis-ratio cut-off stricter ($>0.6$), which removes $50\%$ of our sample, actually improves the negative correlation (the coefficients become more negative by $\sim 40\%$ in Fig.~\ref{fig:bdr_frag_tot}). We further check if this correlation is rather driven by a dependence on the total stellar-mass of the galaxy. While the bulge-to-disk ratio is found to be associated with the stellar-mass, the correlation is much weaker ($r\sim 0.14$) than that with the clumpiness. We therefore claim that this is possibly a second-order effect. We also do not find any detectable correlation between clumpiness and stellar-mass ($r\sim 0.05$).  

Finally, we investigate a dependence on integrated star-formation over the whole galaxy using the star-formation rate (SFR) measurements from spectral fitting in S17. We compare them to the expected  SFR for each galaxy had it been exactly on the star-forming main-sequence \citep{leslie20}. We re-create Fig.~\ref{fig:bdr_frag_tot}, with only the galaxies having measured SFR in S17 using a constant star-formation history model and with values within or above a 0.3 dex scatter of the main-sequence. This filtering is to ensure only confirmed star-forming galaxies are included. The results show no change in the observed correlation (r = $-0.52 \pm 0.08$), and no dependence is observed with the distance of the galaxy from the main sequence (the ratio of SFR observed and that estimated based on the main-sequence). The corresponding figure in shown in the Appendix (Fig.~\ref{fig:bdr_frag_sfr}).

Although we primarily discuss the bulge-to-disk ratio here, replacing it with bulge-to-total ratio maintains the negative correlation, although saturating at $\sim 1$ as expected. When we check the same for the disk-to-total ratio, we observe that there is a very weak positive correlation ($r \sim 0.10$). Both these results suggest that the strong relation observed in Fig.~\ref{fig:bdr_frag_tot} is mainly driven by the relation between the bulge dominance and clumpiness, with weak contributions from a parallel decline of the disk strength. 

\subsection{Bulge-to-disk and clumpiness optical/near-IR ratios}


We further investigate the ratio of the clumpiness   detected in rest-frame optical (F115W) and near-IR (F356W) as a tracer of an equivalent integrated color of the clumps  observed in a galaxy. This value will be associated, but not identical to the color of individual clumps.  

\begin{figure}[!ht]
    \centering
    \includegraphics[width=0.45\textwidth]{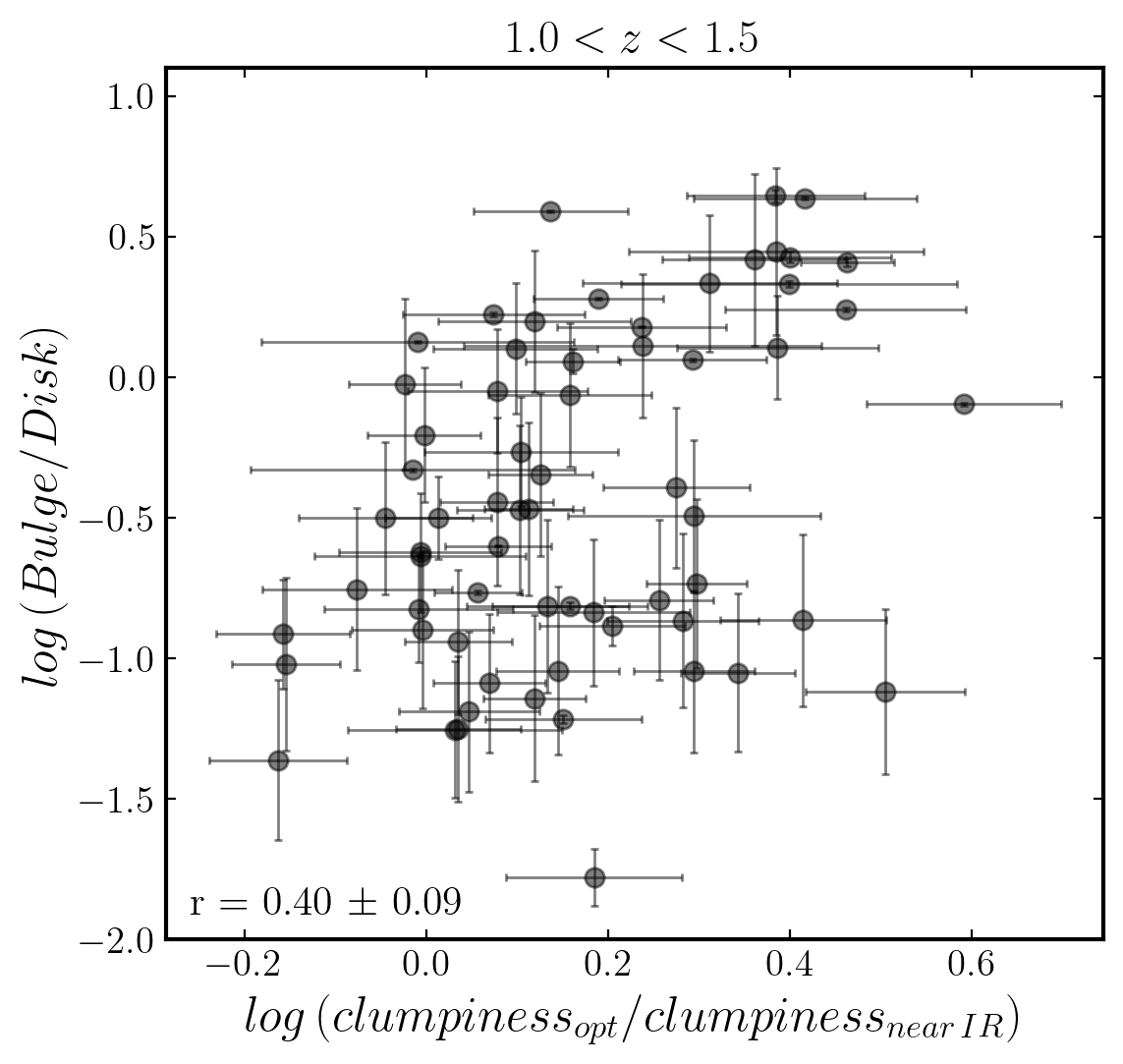}
    \caption{The bulge-to-disk ratio vs the ratio of clumpiness measured in rest-frame optical (f115W) and near-IR (f356w) for redshift $1.0-1.5$.}
    \label{fig:frag_ratio}
\end{figure}

We limit this to the $1.0 < z < 1.5$ range to maximise the wavelength difference between the filters, while ensuring that F115W still traces the optical light and F356W probes the stellar-mass tracing near-IR flux. The latter would not be true at $z > 1.5$. We observe that the clumpiness ratio shows a moderate correlation with the bulge-to-disk ratio ($r = 0.40 \pm 0.09$, Fig.~\ref{fig:frag_ratio}). In other words, for a specific clumpiness in near-IR light sensitive to stellar-mass, those showing higher clumpiness in optical wavelengths show higher bulge-to-disk ratio and vice-versa. However, it should be also be noted that majority of our sample has a clumpiness ratio $> 1$, suggesting that most galaxies have more prominent clumps  in shorter bands tracing younger stars and star-formation. We provide an interpretation of this correlation in Sec.~\ref{sec:sims_int}.

\section{Discussion} \label{sec:conc}
\subsection{Clumpy galaxy fractions}

Our work showcases that clumpiness in high redshift ($z > 1$) galaxies is present across rest-frame UV to near-IR. We also find that there is $ 60 \pm 20\,\%$ spatial overlap between the clumpy maps between these wavelengths. Especially given their presence in the stellar-mass tracing near-IR, we can conclude that these features do play a role in the morphological evolution of galaxies. It is therefore unlikely that we are only dealing with young short lived star-forming objects termed as UV-clumps, demonstrating the need of a more generalised classification of clumps.  This also needs to accommodate the possibility of the clumps  being associated with early stages of structures like spiral arms as suggested by the intensity maps in Fig.~\ref{fig:disk-fragment_detection} and \ref{fig:disk-fragment_detection_examples} \citep[corroborated by detection of `clumpy spirals' at $z > 1$, e.g.,][]{elmegreen09,margalef-bentabol22}. 

We first discuss possible mechanisms of the formation of these structures in lieu of the fraction of galaxies featuring non-zero clumpiness. Our $\gtrsim 40 \%$ fractions are much higher than the fraction of galaxies expected to be undergoing major-mergers. The gas-rich major-merger fraction is a factor of 2 lower based on expectations from \cite{lopez-sanjuan13}. Similarly, the same fraction from \cite{lotz11} is found to be lower (by a factor $> 2 $ up to $z = 1.5$) for merger observability timescales of $\leq 2\, \rm Gyr$ and only comparable if the effects of mergers are observable over a $\sim 3\,\rm Gyr$ interval. We also compare to major-merger fractions from TNG50 and TNG100 simulations \citep{nelson19} and find them to also be a factor $>1.5$ lower that the our clumpy   galaxy fraction, across the stellar-mass range and up to a merger detectability timescale of $1.5\,\rm Gyr$. 

However, the fractions based on minor-mergers \citep{lotz11} and violent disk instabilites \citep{cacciato12} do agree with our estimation of galaxies showing clumpiness   \citep[for expectations from each scenario, check ][]{guo15}. This suggests that a large fraction of the clumpy   galaxies are likely experiencing violent disk instabilities  and/or minor-mergers.

\subsection{Physical interpretation based on simulations} \label{sec:sims_int}

The negative correlation observed between the clumpiness and bulge-to-disk flux ratio (Fig.~\ref{fig:bdr_frag_tot}) may indicate an underlying physical connection. This trend is especially important since we expect this ratio to closely trace the associated mass-ratio due to a minimally varying mass-to-light ratio (Sec.~\ref{subsec: stell_morph}). Furthermore, given that the bulges of galaxies are known to be redder than the disks, either due to high dust obscuration or older stellar populations, the negative correlation would actually get stronger if we considered higher than expected mass-to-light ratio variations.

We now compare this to theoretical expectations to suggest a physical interpretation of this negative correlation. Multiple simulation studies have claimed clumps  in disk galaxies enable the funneling of gas to the centre to form the galactic bulge \citep[see][for a review]{bournaud16}. Our results feature the first observational sample indicating how such a link could manifest. The observed trend in Fig.~\ref{fig:bdr_frag_1} is consistent with an evolutionary trajectory beginning with the appearance of clumps, which leads to the dynamical friction and torques that drive gas (possibly also some the clumps  themselves) to the centre.  
As mentioned in Sec.~\ref{sec:bdr_frag_corr}, there is only a mild contribution to the negative correlation from a decline of the galaxy disks in our sample. Hence we expect the disk to mainly survive this possible scenario. Therefore the clumpiness decreasing with increase in bulge-to-disk ratio would be under the combined effects of migration as well as destruction by stellar feedback \citep{elmegreen08, ceverino10, bournaud11, bournaud14, mandelker14, mandelker17}.

On the other hand, our results also agree with the scenario of bulges leading to the stabilization of the gas in disks \citep{martig09, agertz09, ceverino10, hopkins23}. This process in turn would prevent the formation of clumps resulting in lower clumpiness in galaxies with high bulge-to-disk ratio. Assuming that galaxies without a dominant bulge would be clumpy, this scenario would independently explain the negative correlation in Fig.~\ref{fig:bdr_frag_tot}. Nevertheless, the dynamical effect of the clumps cannot be disregarded. Hence, even if we reject the possibility of the individual clumps migrating to the core, one could still expect the first scenario to be contributing simply through driving gas inwards. This is especially true given that we detect these clumps in the stellar-mass tracing near-IR which will inevitably add to the dynamical friction experienced by the gas within the disks \citep[e.g.,][]{bournaud14}.


Finally, Fig.~\ref{fig:frag_ratio} suggests that the underlying negative slope of the correlation in Fig.~\ref{fig:bdr_frag_1} increases as we go from rest-frame near-IR to optical. It should be noted that this is observed only if one combines the whole sample at $1.0 < z < 1.5$ with galaxies showing clumps in both F115W and F356W filters. It is not as evident separately in the two mass bins shown in Fig.~\ref{fig:8} and \ref{fig:9} in the appendix. This correlation can be interpreted as the ratio of color of clumps and the color of the underlying disks, since the latter is the denominator in the calculation of clumpiness. Therefore, Fig.~\ref{fig:frag_ratio} suggests that galaxies with lower bulge-to-disk ratio have clumps and their parent disks with similar colors. However those with more dominant bulges have clumps that are bluer than their disks.
This could be due to contributions from two separate populations resulting from in-situ as well as ex-situ (accreted clumps/minor-mergers) formation \citep{mandelker14, mandelker17, zanella19}. It is expected that in-situ clumps are younger than their parent galaxies, whereas clumps that fall into the galaxy may introduce older stellar populations. However, this interpretation ignores contributions from dust attenuation resulting in reddening. A definitive conclusion requires studying the spectral energy distribution of individual clumps and placing them within the framework of this study.


\section{Conclusions}

Our investigation aims to simultaneously map clumps  in a mass-complete sample of galaxies at $z=1-2$ along with their underlying stellar morphology. It has been made possible by the high-resolution rest-frame optical and near-IR capabilities of JWST/NIRCam. We find clumps  to not only be limited to optical wavelengths sensitive to young stellar populations (as suggested previously by detection of UV-bright clumps), but also almost equally in near-IR light suggesting an imprint on the stellar distribution. We find a statistically significant correlation between bulge-to-disk ratio, determined in rest frame near-IR, and the clumpiness in individual galaxies. We find a moderate to strong negative correlation between the two: as the clumpiness   decreases, the bulge becomes more prominent. This result strongly suggests that the central bulge is evolutionarily linked to clumps.   Finally, we find that this correlation is steeper in optical than in near-IR, suggesting multiple formation mechanisms of the observed clumps.

\begin{acknowledgments}
We would like to express our gratitude to the anonymous referee who made valuable comments that considerably improved the quality of this work. LCH was supported by the National Science Foundation of China (11721303, 11991052, 12011540375, 12233001), the National Key R\&D Program of China (2022YFF0503401), and the China Manned Space Project (CMS-CSST-2021-A04, CMS-CSST-2021-A06). XD is supported by JSPS KAKENHI Grant Number JP22K14071. BSK would like to thank Benjamin Magnelli and Carlos G\'omez-Guijarro for their brilliant insights into this work and valuable suggestions.  
\end{acknowledgments}



\bibliography{main}{}
\bibliographystyle{aasjournal}



\appendix

\begin{figure*}[!ht]
    \centering
    \includegraphics[width=0.8\textwidth]{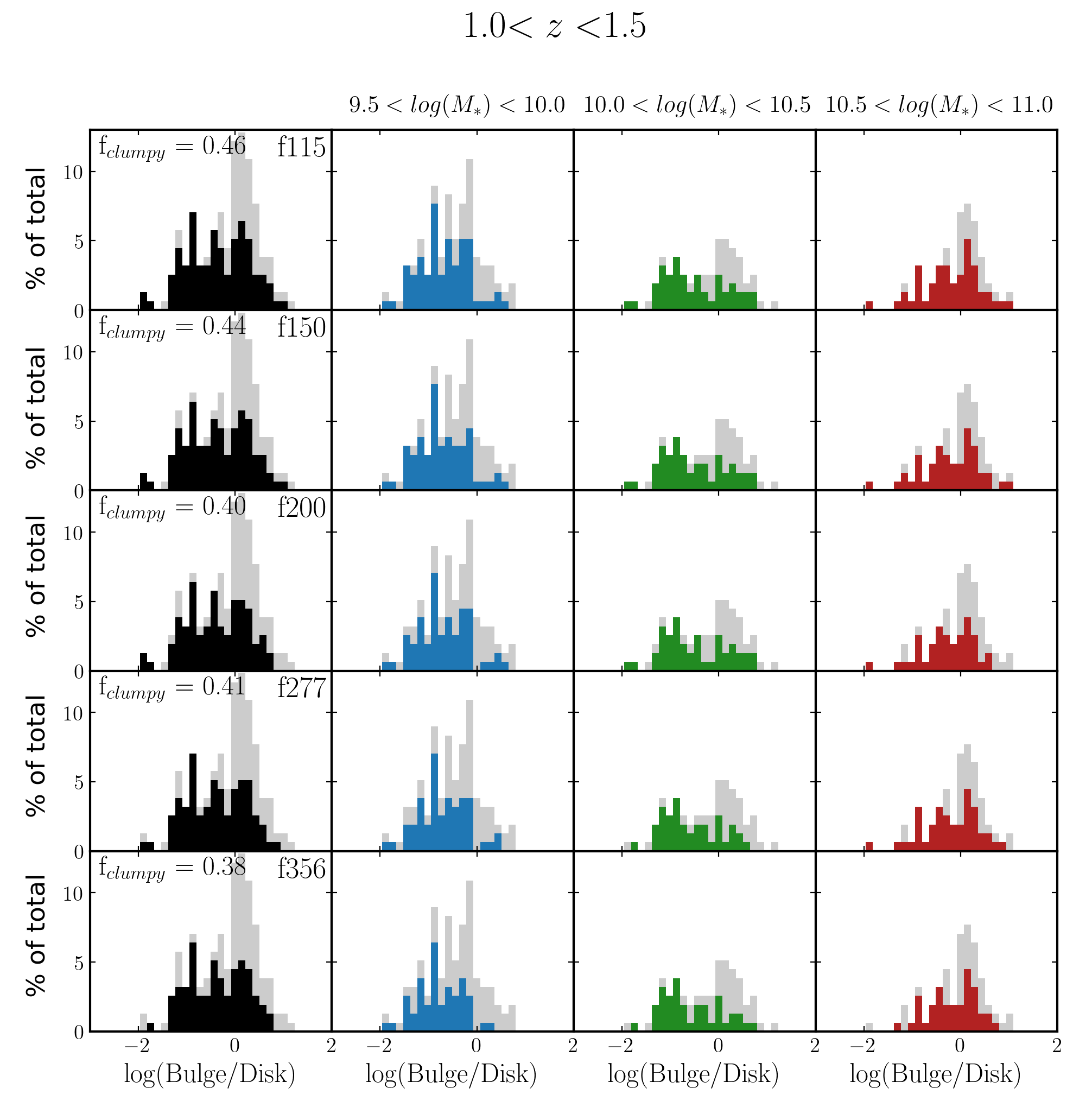}
    \caption{Expansion of Fig.~\ref{fig:hist_1} for $1.0 < z < 1.5$. Please note the addition of the mass-bin $log(M_{*}/M_{\odot}) = 9.5 - 10.0$, which has not been added in the primary work for consistency with the completeness limit at $z > 1.5$. Nevertheless, S17 has a $90\,\%$ mass-completeness at $log(M_{*}/M_{\odot}) = 9.5$ at $1.0 < z < 1.5$.}
\end{figure*}

\begin{figure*}[!ht]
    \centering
    \includegraphics[width=0.8\textwidth]{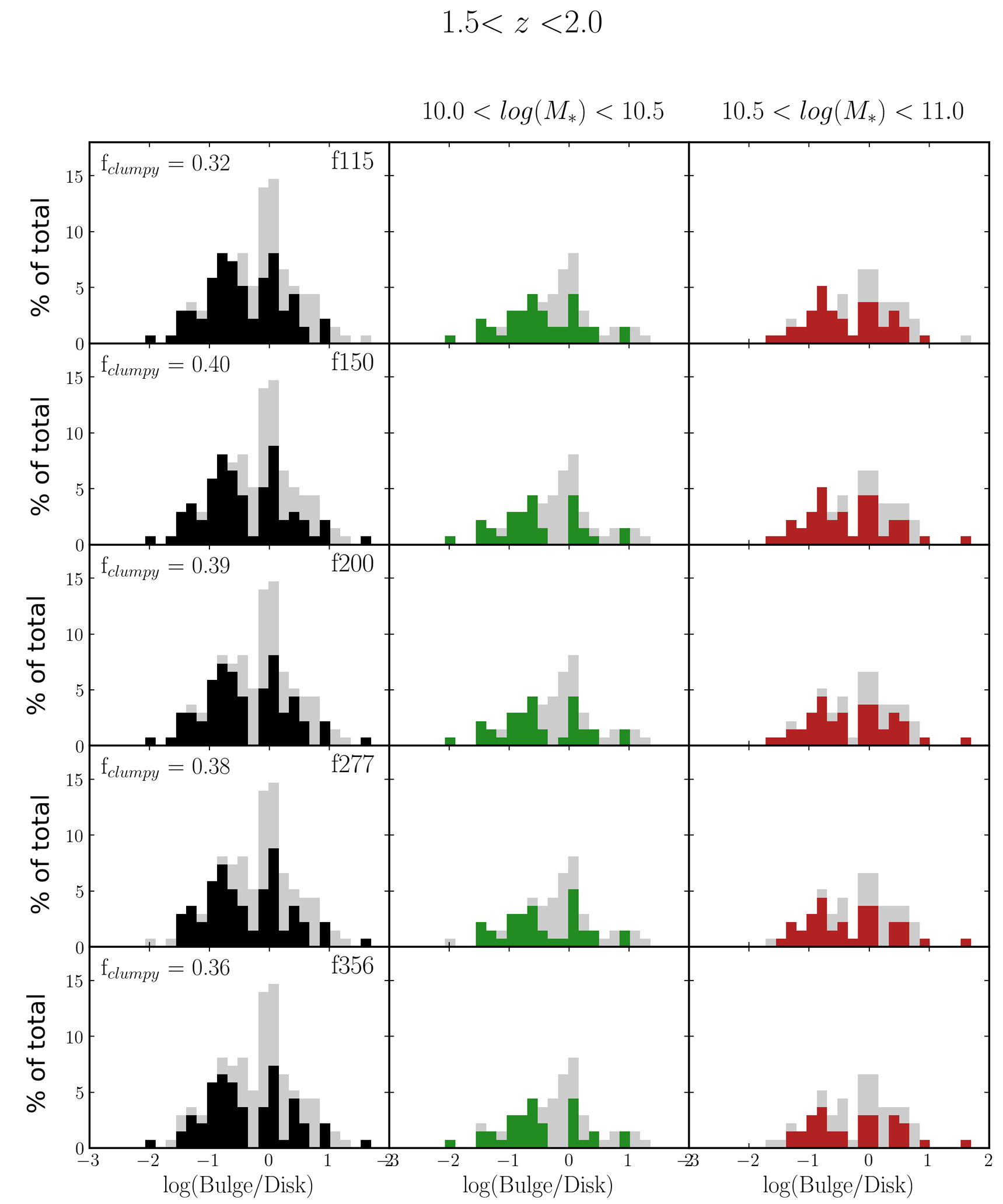}
    \caption{Expansion of Fig.~\ref{fig:hist_1} for $1.5 < z < 2.0$}
\end{figure*}

\begin{figure*}[!ht]
    \centering
    \includegraphics[width=\textwidth]{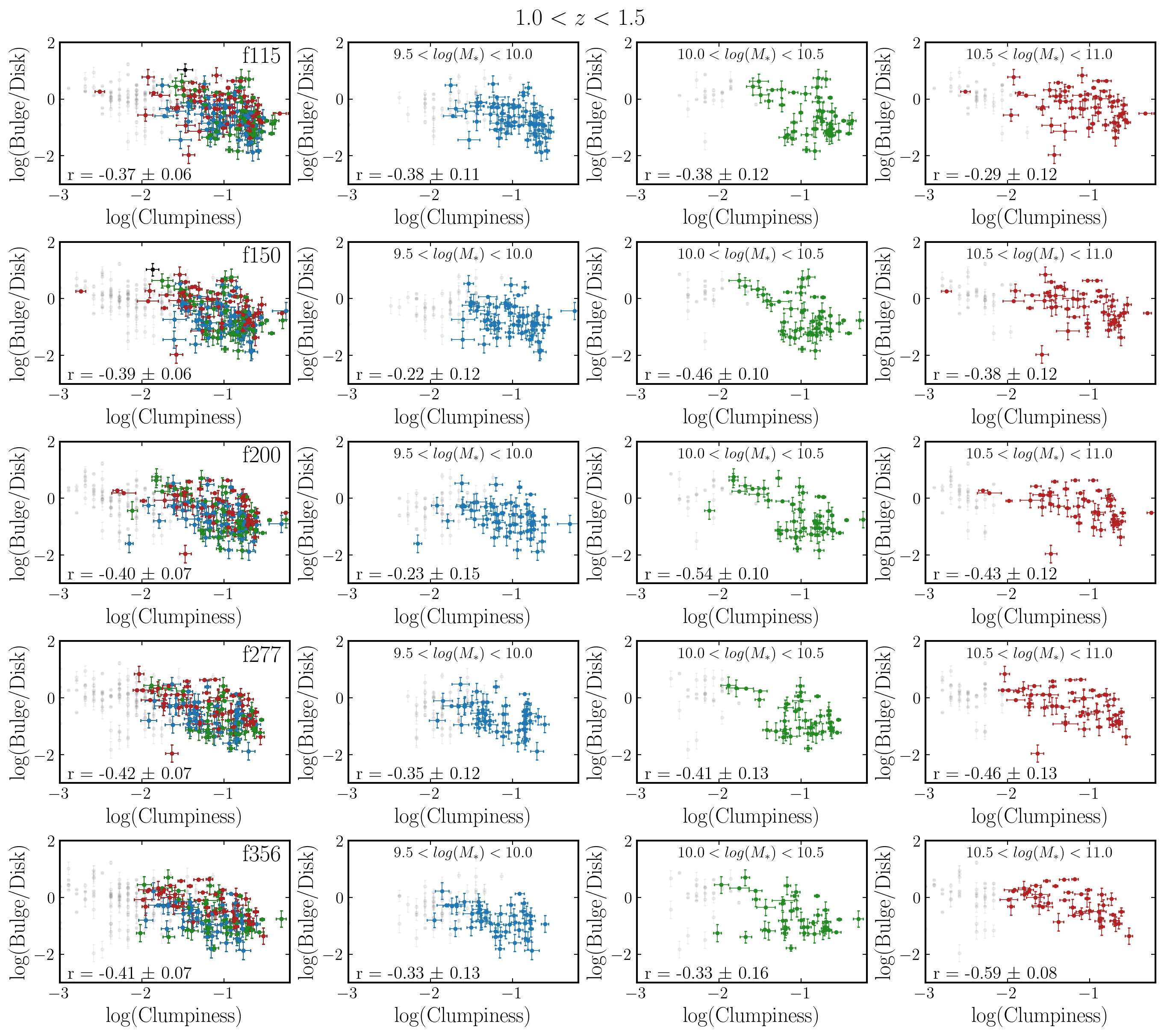}
    \caption{Expansion of Fig.~\ref{fig:bdr_frag_1} for $1.0 < z < 1.5$. Please note the addition of the mass-bin $log(M_{*}/M_{\odot}) = 9.5 - 10.0$, which has not been added in the primary work for consistency with the completeness limit at $z > 1.5$. Nevertheless, S17 has a $90\,\%$ mass-completeness at $log(M_{*}/M_{\odot}) = 9.5$ at $1.0 < z < 1.5$.} \label{fig:8}
\end{figure*}

\begin{figure*}[!ht]
    \centering
    \includegraphics[width=0.8\textwidth]{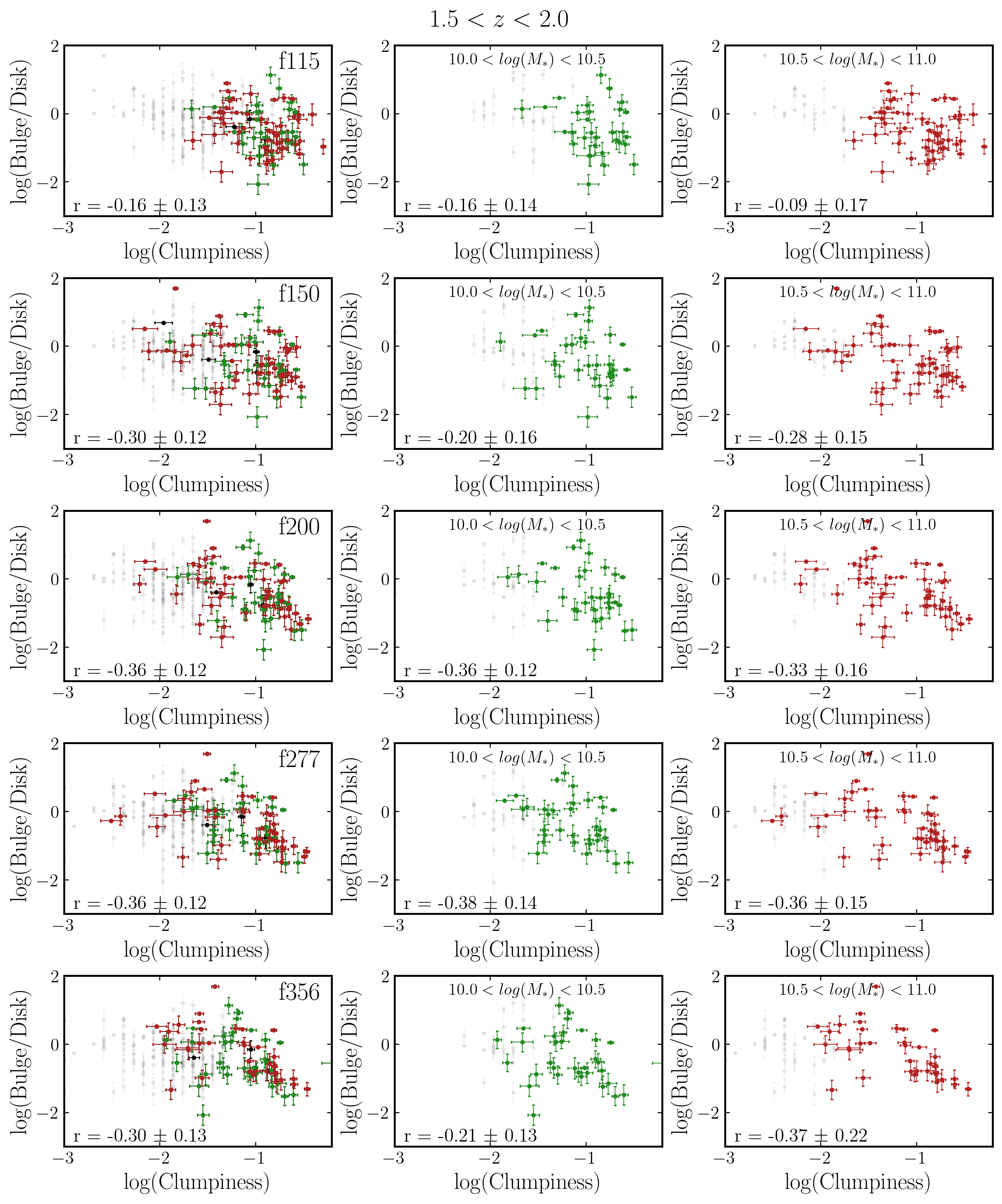}
    \caption{Expansion of Fig.~\ref{fig:bdr_frag_1} for $1.5 < z < 2.0$}\label{fig:9}
\end{figure*}

\begin{figure*}[!ht]
    \centering
    \includegraphics[width=0.8\textwidth]{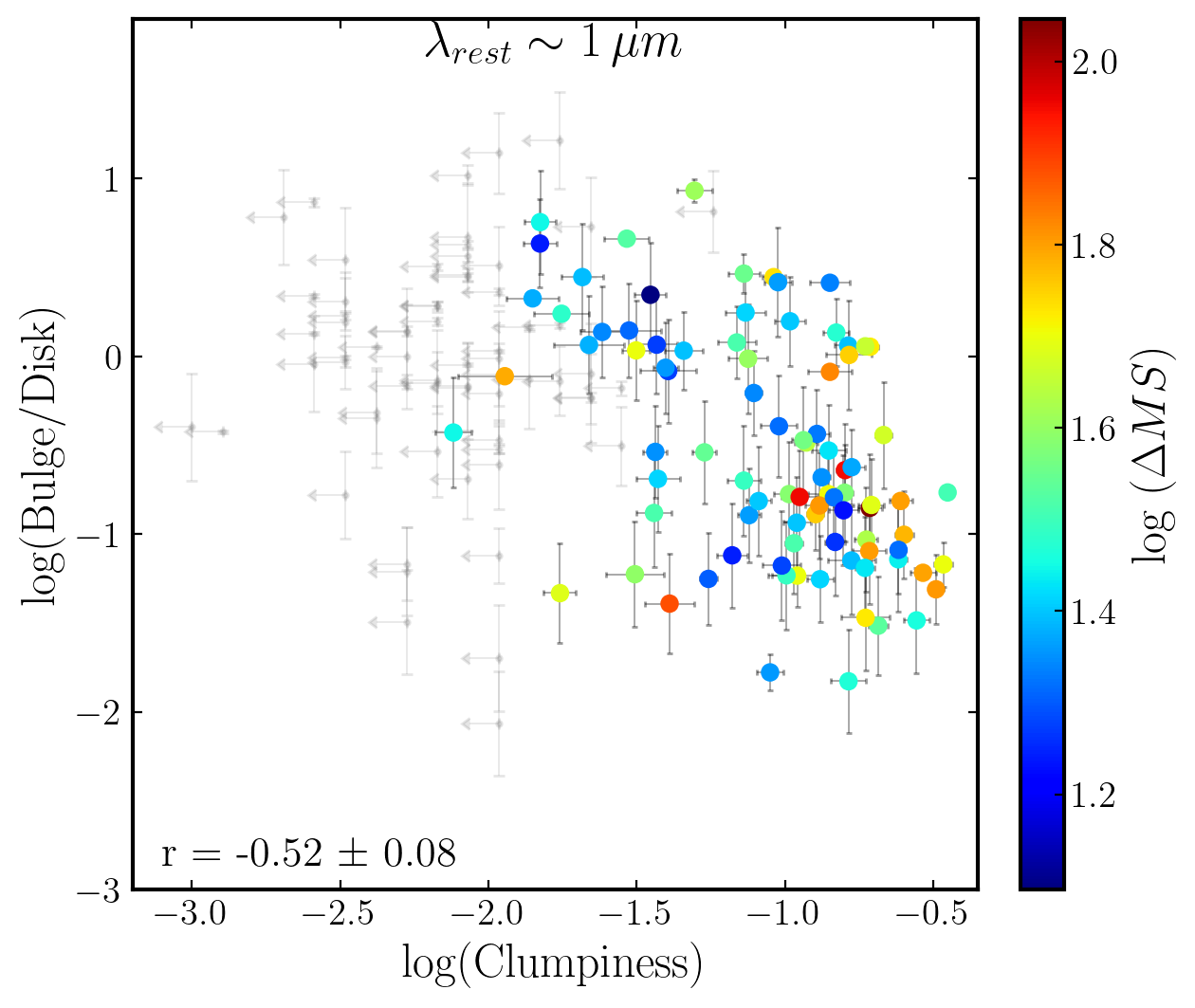}
    \caption{Bulge-to-disk ratio vs clumpiness at rest-frame wavelength $\sim 1\,\mu$m from Fig.\ref{fig:bdr_frag_tot}. Only the galaxies with measured SFR in S17 and with values within or above a 0.3 dex scatter of the star-forming galaxy main sequence \citep{leslie20} for the respective stellar mass and redshift of the galaxies. The colorbar represents the distance from the main sequence, defined as the log of the ratio between the measured SFR and that expected for the galaxy had it been exactly on the main sequence.}
    \label{fig:bdr_frag_sfr}
\end{figure*}

\end{document}